\documentclass[aps,pra,epsfigure,twocolumn,showpacs]{revtex4}
\usepackage{dcolumn}    
\usepackage{bm} 
\usepackage{graphicx}
\usepackage{amsmath}    
\usepackage{latexsym}
\usepackage{amsfonts}   
\usepackage{amssymb}
\usepackage{array}      
\usepackage{epsfig}
\usepackage{times}
\usepackage{txfonts}
\usepackage{epstopdf}
\usepackage{pifont}
\usepackage{dsfont}
\usepackage{amscd}
\usepackage{amsfonts}
\usepackage{textcomp}

\newcommand{\ket}[1]{\left\vert#1\right\rangle}

\newcommand{\bra}[1]{\left\langle#1\right\vert}

\newcommand{\nbar}{\overline{n}}

\newcommand{\be}{\begin{equation}}
\newcommand{\ee}{\end{equation}}
\newcommand{\bea}{\begin{eqnarray}}
\newcommand{\eea}{\end{eqnarray}}

\begin{document}
\title{Enhancing non-classicality in mechanical systems}
\author{Jie Li$^1$, Simon Gr\"oblacher$^{2}$ and Mauro  Paternostro$^{1,3}$}
\affiliation{$^1$Centre for Theoretical Atomic, Molecular and Optical Physics, School of Mathematics and Physics, Queen's University, Belfast BT7 1NN, United Kingdom\\
$^2$Institute for Quantum Information and Matter, California Institute of Technology, 1200 E. California Blvd., Pasadena, CA 91125, USA\\
$^3$Institut f\"ur Theoretische Physik, Albert-Einstein-Allee 11, Universit\"{a}t Ulm, D-89069 Ulm, Germany}

\begin{abstract}
We study the effects of post-selection measurements on both the non-classicality of the state of a mechanical oscillator and the entanglement between two mechanical systems that are part of a distributed optomechanical network. We address the cases of both Gaussian and non-Gaussian measurements, identifying in what cases simple photon-counting and Geiger-like measurements are effective in distilling a strongly non-classical mechanical state and enhancing the purely mechanical entanglement between two elements of the network.
\end{abstract}
\date{\today}
\maketitle

Optomechanics is at the heart of an intense research activity aiming to demonstrate quantum control at large scales and under unfavorable working conditions~\cite{optoreview}. The possibility to exploit massive mechanical systems for the preparation of quantum states and the achievement of non-classical features has inspired a considerable amount of research, including schemes for the generation of fully mechanical entangled states~\cite{allmechanical,Mazzola} and quantum correlations in hybrid systems involving light, mechanical and collective atomic modes~\cite{ibrido}. The flexibility and high potential for hybridization of such systems could be key elements in the design of a new paradigm for devices that are able to perform long-distance quantum communication and networking~\cite{kimble,paternostrovari}. Indeed, the design of a quantum network consisting of remote optomechanical systems connected by optical fields is not only foreseeable but also the subject of ongoing theoretical investigations~\cite{braunstein,Mazzola,borkje,aoki}. 

The realization of such a vision requires effective strategies for the induction of non-classical features at both the single- and multi-mode mechanical level. That is, we devise techniques for  getting non-classicality in single-mode mechanical states (for quantum state engineering, for instance) and distribute quantum correlations among remote sites of a network. Both these goals have recently received attention~\cite{Mazzola,Paternostro2012,HuangAgarwal,HuangAgarwal2}, where Huang and Agarwal~\cite{HuangAgarwal}, for instance, have shown that by injecting an optomechanical cavity with squeezed light, a squeezed state of the mechanical mode can be obtained. Other approaches, based on more sophisticated optomechanical Hamiltonians, have shown the possibility of squeezing a mechanical oscillator and enhancing optomechanical entanglement through time-modulation~\cite{timemodulation}, while intrinsic nonlinear mechanisms have been used in Ref.~\cite{hartmann} to generate phononic Fock states. 

In this paper, we take a complementary approach and show that, by incorporating the non-linear effects induced by measurement post-selection into the design of an optomechanical network, one can benefit in two ways. First, there would be no need for the use of non-classical states of light to induce strongly non-classical states of a mechanical mode. Second, purely mechanical entanglement can be generated, well above the performances of early suggestions in this respect~\cite{Mazzola}, and robustly with respect to the effects of the surrounding environment or unfavorable working points. We determine the type of measurements needed to achieve such tasks, compare relative performances and investigate their resilience to noise. Our study is a step forward in the design of experimentally implementable schemes for the achievement of non-classicality in a quantum network of genuinely mesoscopic quantum nodes and is fully in line with other proposed means to induce  non-classicality in mechanical systems at the quantum level, such as the methods put forward in Ref.~\cite{Jiang,Yanbei}.

The paper is organized as follows: In Sec.~\ref{nonclassicality} we introduce our model and provide an intuitive picture for the achievement of non-classicality in the state of an optically driven mechanical mode. In Sec.~\ref{nonclass} we present our results for the experimentally realistic case of a strongly pumped open-cavity system, showing that mechanical-mode states with negative associated Wigner function can be obtained by simple post-selection strategies. Sec.~\ref{ent} is devoted to the extension of our approach to a two-cavity system that embodies the smallest cell of an optomechanical quantum network. Finally, in Sec.~\ref{conc} we present our conclusions. We include an Appendix where we give details of our approach that, albeit relevant and useful for the reader, are not central to our discussion.

\begin{figure}
\includegraphics[width=0.6\linewidth]{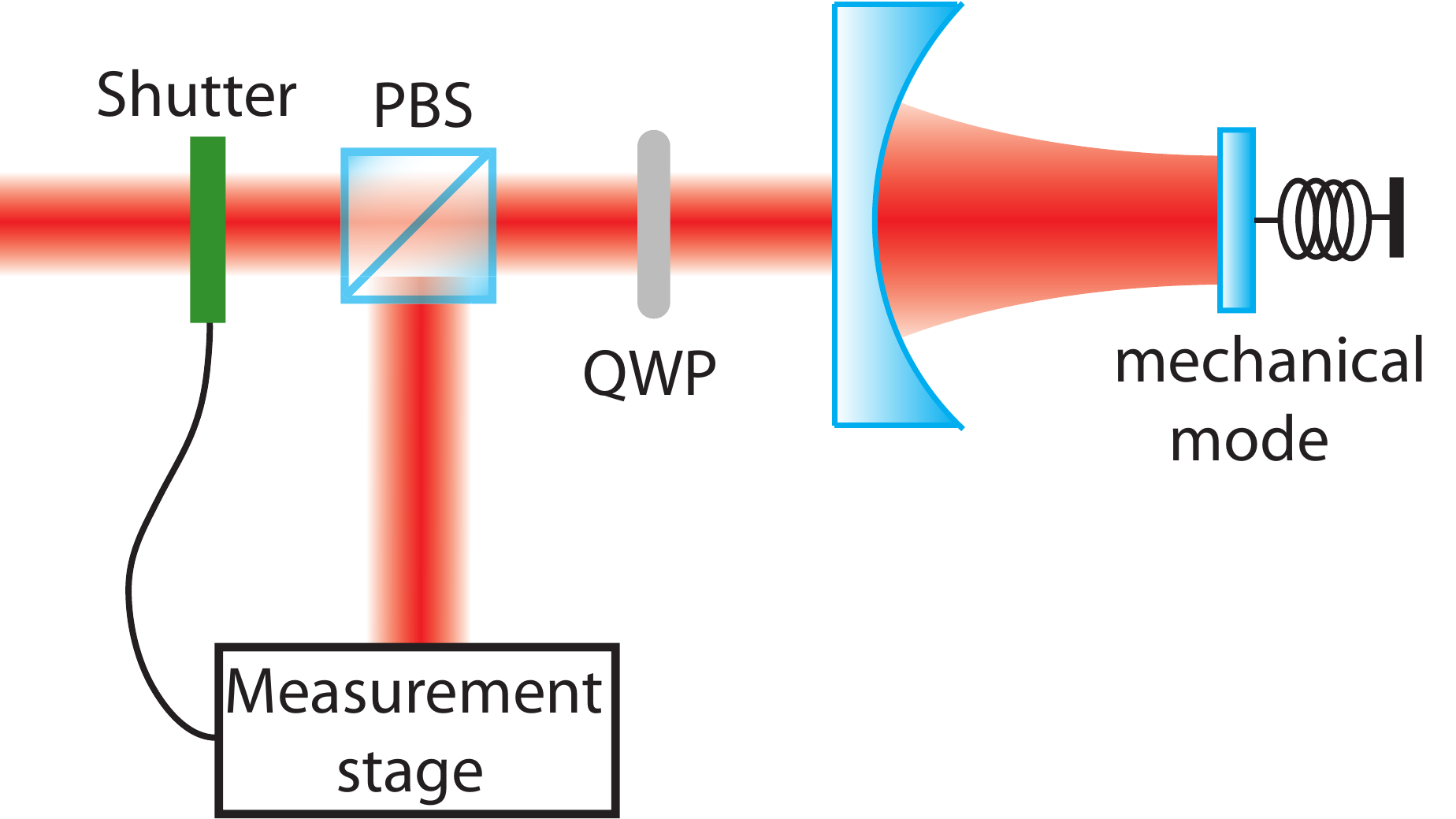}
\caption{Sketch of the proposed experimental setup. A horizontally polarized pump field at frequency $\omega_p$ drives a cavity with a vibrating end mirror of mass $m$ and mechanical frequency $\omega_m$. The field leaking out of the cavity enters a measurement stage. The optical routing occurs via a double-passage through a quarter wave plate (QWP) that changes the polarization of the beam to vertical and a polarizing beam splitter (PBS) that transmits (reflects) horizontally (vertically) polarized light. As the optical signal enters the measurement stage, a shutter is electronically activated to block the pump field.}
\label{modelsingle}
\end{figure}

\section{The model and a simplified picture}
\label{nonclassicality}

We start analyzing the system composed of a single optomechanical device with the objective of identifying conditional strategies for the achievement of non-classical mechanical states.  In order to provide a complement to the situation that will be addressed later on in this paper and that is very close to the current experimental possibilities, we resort to a picture where all sources of damping and dephasing are neglected as in the case of strong single-photon radiation-pressure coupling considered recently in a series of exploratory theoretical works~\cite{Nunnenkamp, He}. In such a unitary picture, assuming resonant light-cavity coupling, the evolution of the system in the laboratory frame is governed by the model 
\be
\label{unitary}
\hat H_{u}=\hbar\omega_c\hat{n}+\hbar\omega_m\hat b^\dag\hat b-\hbar\chi\hat{n}\hat{q},
\ee
where $\omega_c$ ($\omega_m$) is the cavity (mechanical) frequency and $\chi$ the optomechanical coupling rate. For instance, in a linear Fabry-Perot cavity this is given by $\chi=\chi_0\sqrt{\hbar/(2m\omega_m)}$, where $\chi_0=\omega_c/L$, $L$ is the length of the cavity and $m$ the effective mass of the mechanical oscillator. However, it should be stressed that the validity of our model and approach is general and does not rely on the specific details of the experimental setting being considered. As such, our techniques can be well applied to the systems used in~\cite{Painter}. We have introduced the mechanical mode position-like quadrature operator $\hat{q}=(\hat b+\hat b^\dag)/\sqrt2$, the cavity photon-number operator $\hat{n}=\hat a^\dag\hat a$ , and  the optical and mechanical annihilation (creation) operators $\hat a$ ($\hat a^\dag$) and $\hat b$ ($\hat b^\dag$). The pump term has been dropped by assuming the cavity field as prepared in a desired state. Mancini {\it et al.}~\cite{Mancini} and Bose {\it et al.}~\cite{Bose} solved the dynamics induced by Eq.~(\ref{unitary}) fully, providing the form of the time propagator 
\be
\label{tp}
\hat{\cal U}(t)=e^{-i{\omega_c}\hat n t}e^{i \frac{\chi^{2}}{\omega^2_m}\hat n^2[\omega_m t-\sin(\omega_mt)]}\hat{\cal D}_m\left[\frac{\chi(1-e^{-i\omega_mt})}{\omega_m}\hat n\right]e^{-i\omega_m\hat b^\dag\hat b t},
\ee
which has later been reprised to show that, as long as such a unitary picture is considered, optomechanical entanglement exists at any finite temperature of the mechanical system~\cite{Ferreira}. Here we assume that the cavity field is prepared in a coherent state $\ket{\alpha}$ and is interacting with a mechanical system that is at thermal equilibrium (at temperature $T$) with its own phononic environment. We thus describe the state of the mechanical system as $\rho_m=\int d^2\beta P_{th}(\beta,\nbar)\ket{\beta}_m\!\bra{\beta}$, where $P_{th}(\beta,\nbar)=\frac{1}{\pi\nbar}e^{-\frac{|\beta|^2}{\nbar}}$ is the Gaussian $P$ function of a thermal state with an average number of thermal excitations $\nbar$~\cite{Walls}. We now use this simplified picture to get an intuition of the mechanism behind the achievement of non-classicality in the state of the mechanical system upon projections performed on the optical mode. Such intuition will then be compared to the situation encountered currently in the laboratory, where the linearized picture described before is very appropriate.  

The state of the mechanical system, after the application of the conditional protocol described before and illustrated in Fig.~\ref{modelsingle}, reads
\be
\rho'_m(t)=\frac{\mathrm{Tr}_c[\hat\Pi_c\rho_{cm}(t)]}{\mathrm{Tr}[\hat\Pi_c\rho_{cm}(t)]}
\ee
with $\hat\Pi_c$ the projector, defined in the Hilbert space of the cavity field, that describes the outcome of the measurements on the optical subsystem and $\rho_{cm}(t)=\hat{\cal U}(t)(\rho_{m}\otimes\ket{\alpha}_c\!\bra{\alpha})\hat{\cal U}^{-1}(t)$ the evolved state of the optomechanical system. Following~\cite{Bose} and taking $\alpha\in\mathbb{R}$ for simplicity,
\be
\hat{\cal U}(t)\ket{\alpha,\beta}_{cm}=e^{-\alpha^2/2}\sum^\infty_{n=0}\frac{\alpha^n}{\sqrt{n!}}e^{i\varphi_n(t)}\ket{n}_c\otimes\ket{\phi_n(t)}_m,
\ee
where we have introduced the phase factor $\varphi_n(t)=\frac{\chi^{2}}{\omega^2_m}n^2[\omega_mt-\sin(\omega_mt)]$ and the coherent states of the mechanical mode $\ket{\phi_n(t)}_m=|{\beta e^{-i\omega_mt}+{\chi}n(1-e^{-i\omega_mt})/{\omega_m}}\rangle_m$. As discussed in Ref.~\cite{Bose}, here photon-counting is ineffective for creating non-classical mechanical states, as it will give us only an incoherent superposition of Gaussian states of the mechanical system. On the other hand,  both homodyne and heterodyne measurements turn out to be quite effective. The first (second) measurement is formally equivalent to projections onto quadrature eigenstates (coherent states) of the optical mode~\cite{Walls}, so that $\hat\Pi^{hom}_c=\ket{x(\theta)}_c\!\bra{x(\theta)}$ ($\hat\Pi^{het}_c=\ket{\sigma}_c\!\bra{\sigma}$), with $\ket{x(\theta)}_c$ the eigenstate of the arbitrary quadrature operator $(e^{-i\theta}\hat a+e^{i\theta}\hat a^\dag)/\sqrt2$ ($\ket{\sigma}_c$ a coherent state). In the following we concentrate on the case of $\theta=0$, i.e. projections onto the in-phase quadratures eigenstates, when dealing with homodyne measurements. 

\begin{figure}[t!]
{\bf (a)}\hskip3.5cm{\bf (b)}\\
\includegraphics[width=0.5\linewidth]{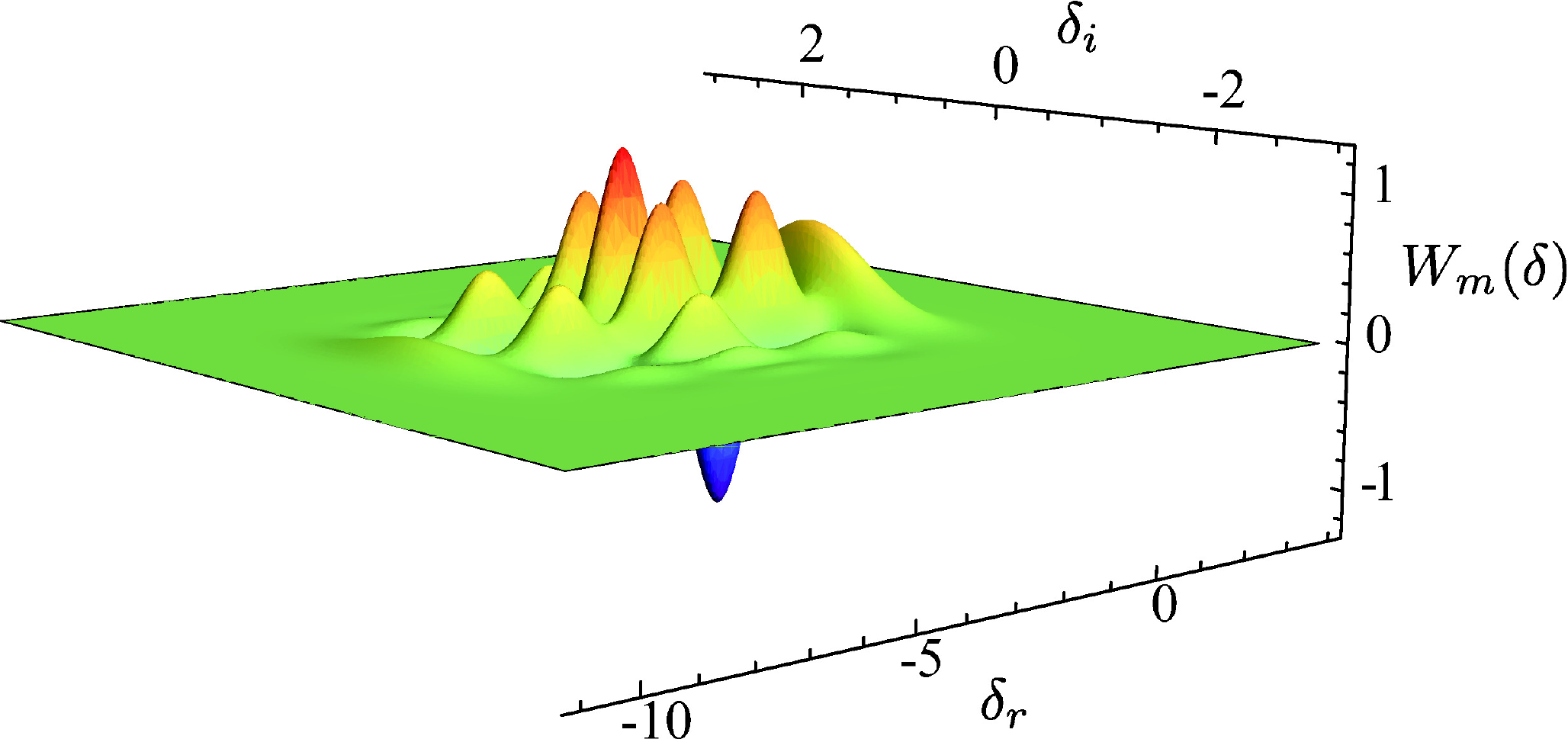}~\includegraphics[width=0.5\linewidth]{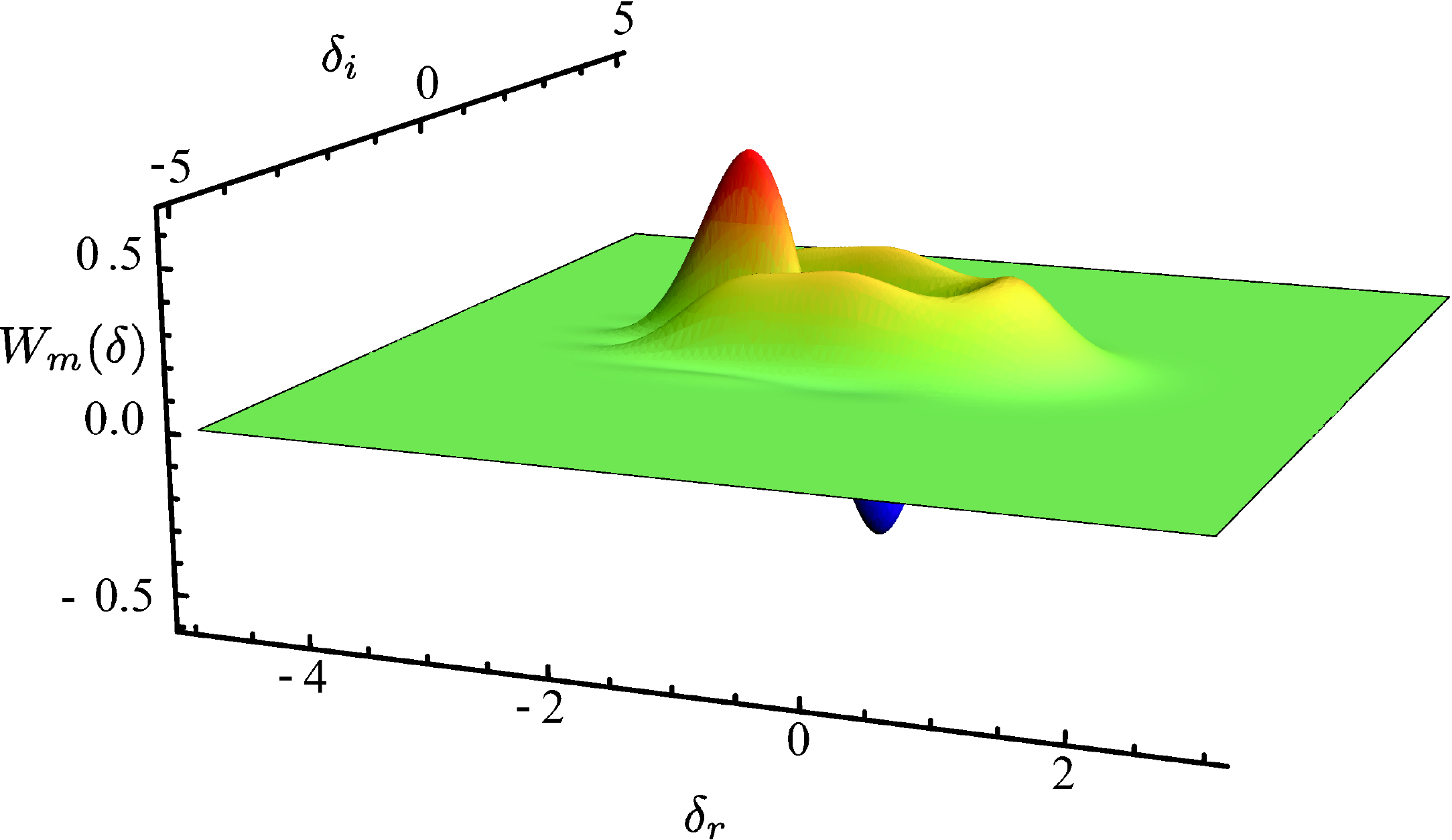}
\caption{{\bf (a)} Wigner function of the mechanical system upon conditional homodyne measurement on the optical mode. We have taken $\nbar=0$, $\chi=\omega_m$, $\alpha=1$ and $t=\pi/\omega_m$, postselecting the measurement outcome $x=0$. {\bf (b)} Same as in panel {\bf (a)} but for a heterodyne measurement on the optical field, postselecting the outcome $\sigma=1$.}
\label{WignerUnitary}
\end{figure}

\begin{figure}[b!]
{\bf (a)}\hskip3.5cm{\bf (b)}\\
\includegraphics[width=0.5\linewidth]{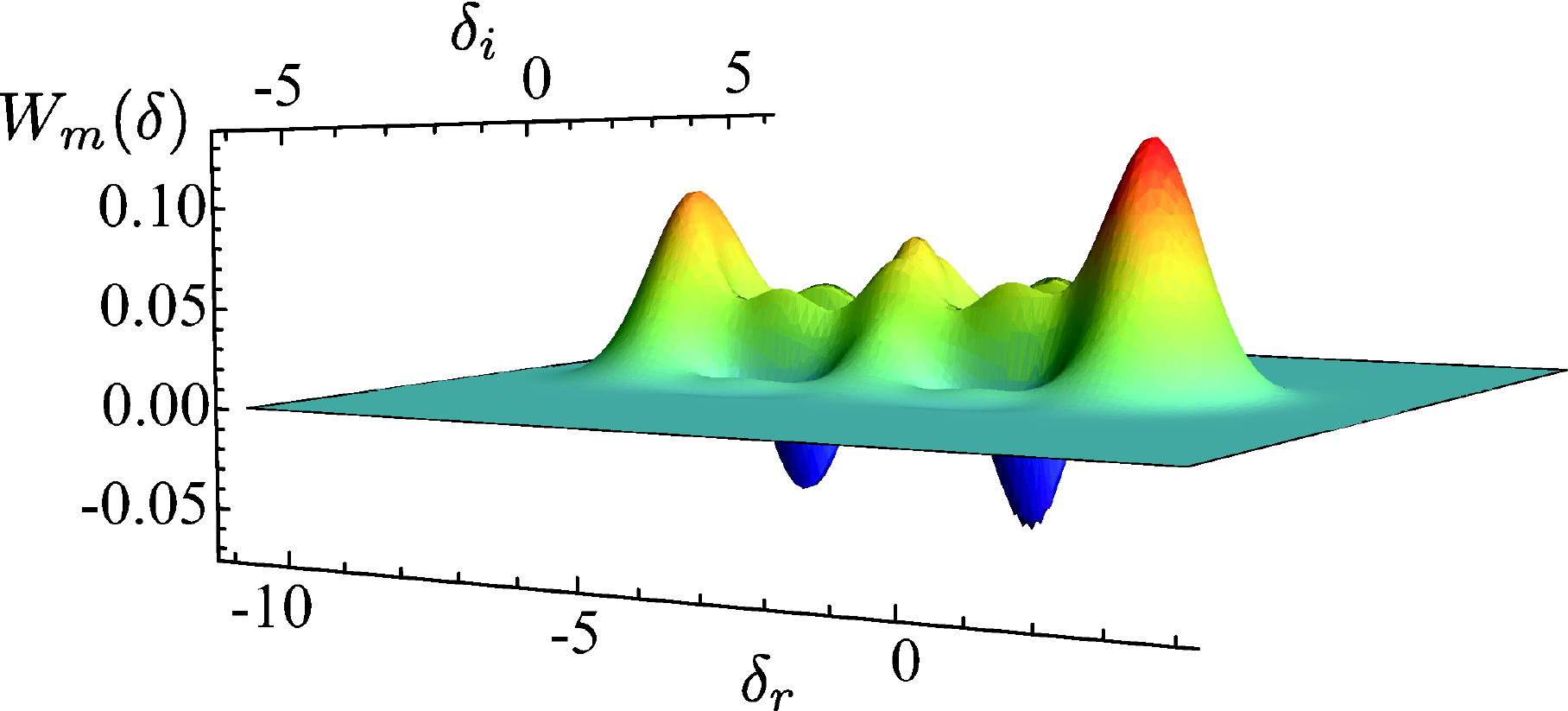}~\includegraphics[width=0.5\linewidth]{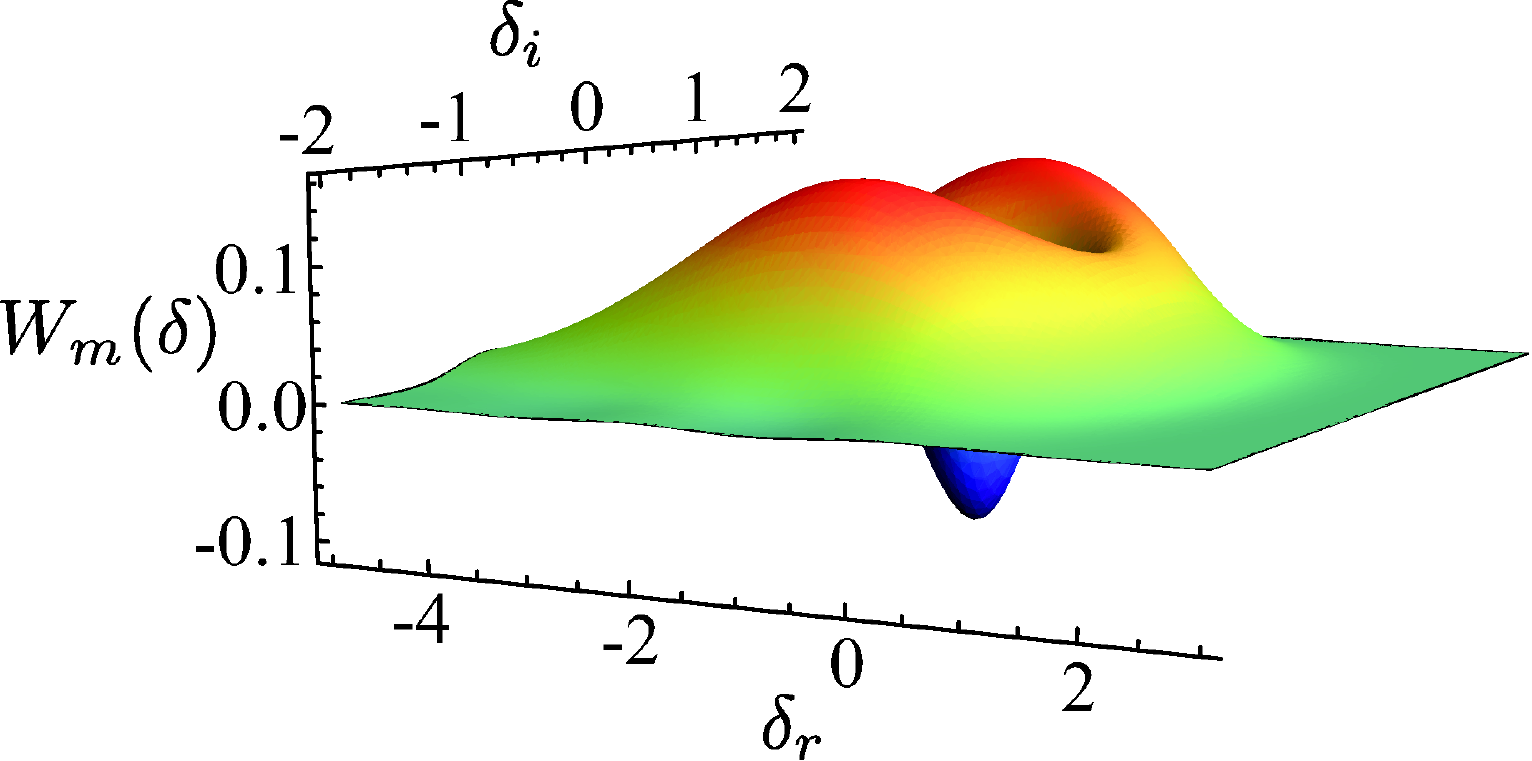}
\caption{{\bf (a)} Wigner function of the mechanical system upon conditional homodyne measurement on the optical mode. We have taken $\nbar=0.5$, $\chi=\omega_m$, $\alpha=1$ and $t=\pi/\omega_m$, postselecting the measurement outcome $x=0$. {\bf (b)} Same as in panel {\bf (a)} but for a heterodyne measurement on the optical field, postselecting the outcome $\sigma=1$.}
\label{WignerUnitaryTemperature}
\end{figure}

In Fig.~\ref{WignerUnitary} we show the Wigner function $W_m(\delta)$ as a function of the values taken by the phase-space variable $\delta=\delta_r+i\delta_i$ of the state of the mechanical mode for both homodyne and heterodyne measurements performed on the field, showing the effectiveness of the measurement postselection process for the purposes of inducing non-classicality in the state of the mechanical mode. In what follows, we adopt the negativity of the Wigner function as the indicator of the non-classicality of the mirror~\cite{Kenfack,Jie}. This is defined as
\begin{equation}
{\cal N_W}=\left|\!\int_{\Phi}W_m(\delta)\text{d}^2\delta\,\right|,
\label{eq13}
\end{equation}
where $\Phi$ is the region of the negative Wigner distribution in phase space.
At proper values of the parameters in Eq.~\eqref{unitary}, the Wigner function takes significantly negative values, therefore unambiguously proving the non-classical nature of the corresponding state, regardless of the measurement performed over the optical subsystem (i.e. both for homodyne and heterodyne measurements). In the case of homodyning, the Wigner function of the mechanical mode resembles that of an unbalanced multi-component Schr\"odinger cat state. The negativity, though, appears to be quite fragile with respect to the thermal character of the initial mechanical state. A small increase of the initial mechanical mean excitation number from $\nbar=0$ to $\nbar=0.5$ shrinks the negative peak shown in the homodyne case by almost two orders of magnitude and by a factor of 5 in the case of heterodyne measurements, which appear to provide results that are more robust to temperature (cf. Fig.~\ref{WignerUnitaryTemperature}). At $\nbar=2$ already, no negativity can be detected, even at the expenses of a larger optomechanical coupling rate, regardless of the measurement scheme considered. As we will see in the next Section, these results are unique of the fully non-linear interaction model in Eq.~(\ref{unitary}): the situation is strikingly different when the linearised picture typical of strongly pumped optomechanical cavities is considered.

\section{Inducing mechanical non-classicality}
\label{nonclass}

We now show that, by strongly pumping the cavity and modifying the post-processing step, robust mechanical non-classicality can be achieved, robustly with respect to environmental effects causing photon losses and phononic dephasing. 


In order to do this, we abandon the unitary picture adopted above and resort to the full open-system dynamics of the device. In particular, we have to incorporate the effects of a driving field pumping the resonator, losses from a leaky cavity with only a finite quality factor, and decoherence affecting the mechanical mode (at thermal equilibrium). In a rotating frame at the frequency $\omega_L$ of the pump field, the Hamiltonian model in Eq.~(\ref{unitary}) is changed into
\begin{equation}
\label{pumped}
\hat{H}=\hbar\Delta\hat{n}-\hbar\chi\hat{n}\hat{q}+\hbar\omega_m(\hat{p}^2+\hat{q}^2)+i\hbar\varepsilon(\hat{c}^{\dag}-\hat{c}),
\end{equation}
where $\hat{p}$ is the dimensionless momentum-like quadrature operator of the mechanical mode, $\hat{c}$ ($\hat{c}^{\dag}$) is the annihilation (creation) operator of the cavity field (whose energy decay rate is $\kappa$), $\Delta=\omega_c-\omega_L$ is the pump-cavity detuning, and $\varepsilon$ the coupling between the external pump and the cavity field (its value is directly related to the intensity of the pumping field). For a linear Fabry-Perot cavity,  we have $\varepsilon=\sqrt{{2\kappa P}/{\hbar\omega_L}}$, with $P$ the power of the pumping field. In the following, we will assume a strongly pumped cavity ({\it i.e.} $\varepsilon/\omega_m\gg{1}$) with a large number of intra-cavity photons. As it is standard in these conditions, one can approximate the elements of the vector of quadrature operators of the mechanical and optical mode $\hat{O}=(\hat{q}, \hat{p}, \hat{x}, \hat{y})^T$  [with $\hat{x}=(\hat{c}^\dag+\hat c)/\sqrt2$, $\hat{y}=i(\hat{c}^\dag-\hat c)/\sqrt2$] as $\hat{O}_i\simeq\langle\hat{O}_i\rangle+\delta\hat O_i$, where $\langle\hat{O}_i\rangle$ is the mean value of each operator and $\delta\hat{O}_i$ the corresponding fluctuation~\cite{vitalivedral,mauroNJP}. In this way, the interaction Hamiltonian between mechanical mode and cavity field takes a linear form that strongly simplifies the analytic approach to the problem. Finally, the substrate onto which the mechanical system is fabricated induces decoherence due to a background of phononic modes at non-zero temperature. In turn, these give rise to mechanical Brownian motion. 
The resulting dynamics of the quantum fluctuation operators is described by the set of Langevin equations
\begin{equation}
\partial_t\delta\hat{O}{=}{\bf K}\delta\hat{O}+\hat{N}
\label{eq2}
\end{equation}
with $\delta\hat{O}=(\delta\hat{q}, \delta\hat{p}, \delta\hat{x}, \delta\hat{y})^T$  
and
\begin{equation}
{\bf K}=
\begin{pmatrix}
0 & \omega_m & 0 & 0 \\
-\omega_m & -\gamma_m & 2\chi\text{Re}[c_s] & 2\chi\text{Im}[c_s]  \\
-2\chi\text{Im}[c_s] & 0 & -\kappa & \tilde\Delta \\
2\chi\text{Re}[c_s] & 0 & -\tilde\Delta & -\kappa \\
\end{pmatrix}.
\label{eq3}
\end{equation}
In Eq.~(\ref{eq3}) we have introduced the mean value of the cavity-field amplitude $\langle\hat{c}\rangle\equiv c_s=\varepsilon/(\kappa+i\tilde\Delta)$, the  steady-state displacement of the mechanical mode $\langle\hat{q}\rangle\equiv q_s={\hbar\chi_0|c_s|^2}/{m\omega_m^{2}}$, and the effective cavity-pump detuning $\tilde\Delta=\omega_c-\omega_L-\chi_0 q_s$. In what follows, we assume $|c_s|\gg{1}$, which allows us to safely neglect second-order terms in the expansion of each $\hat O_i$ and thus makes the linearised approach rigorous. The values of the parameters assumed in the rest of our work fully guarantee the validity of such assumption and approach. The last term in Eq.~\eqref{eq2} is the vector of zero-mean input noise $\hat{N}=(0,\hat{\xi}{/}\sqrt{\hbar m\omega_m}, \sqrt{2\kappa}\delta\hat{x}_{in}, \sqrt{2\kappa}\delta\hat{y}_{in})^T$, where $\hat{\xi}$ is the Langevin force operator accounting for the Brownian motion affecting the mechanical mode. For large mechanical quality factors, the dimensionless operator $\hat{\zeta}=\hat{\xi}{/}\sqrt{\hbar m\omega_m}$ is auto-correlated as $\langle\hat{\zeta}(t)\hat{\zeta}(t')\rangle{=}\frac{\gamma_m}{2\pi\omega_m}\int\omega \text{e}^{-i\omega(t-t')}[1+\text{coth}(\frac{\hbar\omega}{2k_BT})]\text{d}\omega$ with $k_B$ the Boltzmann constant, $T$ the temperature of the phononic environment, and $\gamma_m$ the damping rate of the mechanical oscillator. Finally, $\delta\hat{x}_{in}=(\delta\hat{c}_{in}^{\dag}+\delta\hat{c}_{in})/\sqrt{2}$, $\delta\hat{y}_{in}=i(\delta\hat{c}_{in}^{\dag}-\delta\hat{c}_{in})/\sqrt{2}$ are the quadratures of the input noise to the cavity.  

The dynamics of the elements of $\delta\hat{O}$ can be solved in the frequency domain by taking the Fourier transform of the Langevin equations~\eqref{eq2}. The corresponding expressions are provided in the Appendix. Any correlation function of pairs of fluctuation operators is then obtained as
\begin{equation}
V_{jk}(t)=\frac{1}{4\pi^2}\iint\!\text{d}\omega\text{d}\Omega e^{-i(\omega+\Omega)t}V_{jk}(\omega,\Omega),
\label{eq4}
\end{equation}
with $V_{jk}(\omega,\Omega)=\langle\{\hat{v}_j(\omega),\hat{v}_k(\Omega)\}\rangle/2 ~(j,k{=}1,..,4)$ the frequency-domain correlation function between elements $j$ and $k$ of $\hat{v}=(\delta\hat{q}, \delta\hat{p}, \delta\hat{x}, \delta\hat{y})$.
The matrix ${\bf V}(t)$ with elements defined as in Eq.~\eqref{eq4} embodies the time-dependent covariance matrix (CM) of the optomechanical system. The linearity of the model treated here guarantees that the dynamical map at hand preserves the Gaussian nature of any input state. 

In order to account for the possibility of feeding the cavity with a non-classical state of light (besides the strong pump that is needed to drive the mechanical mode and perform the linearization), in what follows we consider the general case of input noise characterized by the correlation vector $C=(\langle\delta\hat{c}_{in}^{\dag}\delta\hat{c}_{in}\rangle, \langle\delta\hat{c}_{in}\delta\hat{c}_{in}^{\dag}\rangle, \langle\delta\hat{c}_{in}\delta\hat{c}_{in}\rangle, \langle\delta\hat{c}_{in}^{\dag}\delta\hat{c}_{in}^{\dag}\rangle)\delta(t-t')=(N, N{+}1, e^{-i\omega_m(t+t')}M, e^{i\omega_m(t+t')}M^*)\delta(t-t')$. Taking $N=\sinh^2{r}$ and $M=\sinh{r}\cosh{r}e^{i\phi}$, we describe the case of an input squeezed vacuum state, a case that is general and interesting enough for our goals to embody a useful benchmark. Using the approach sketched above, the CM of the system can be fully determined and, using this tool, one can calculate explicitly the 
characteristic function of the system as $\chi(\alpha, \beta){=}\exp[-O^T {\bf V}(t) O]$. In turn this gives us access to the density matrix of the optomechanical device as~\cite{CahillGlauber}
\begin{equation}
\rho_{cm}=\frac{1}{\pi^2}\iint \text{d}^2\alpha\,\text{d}^2\beta\,\chi(\alpha, \beta)\hat{D}_m(-\alpha)\hat{D}_c(-\beta),
\label{eq11}
\end{equation}
where $\hat{D}_j(\mu)$ is the Weyl operator (amplitude $\mu\in\mathbb{C}$) of mode $j=m,c$~\cite{Walls}.

\begin{figure*}[t]
{\bf (a)}\hskip4.7cm{\bf (b)}\hskip4.7cm{\bf (c)}
\includegraphics[width=1.2\columnwidth]{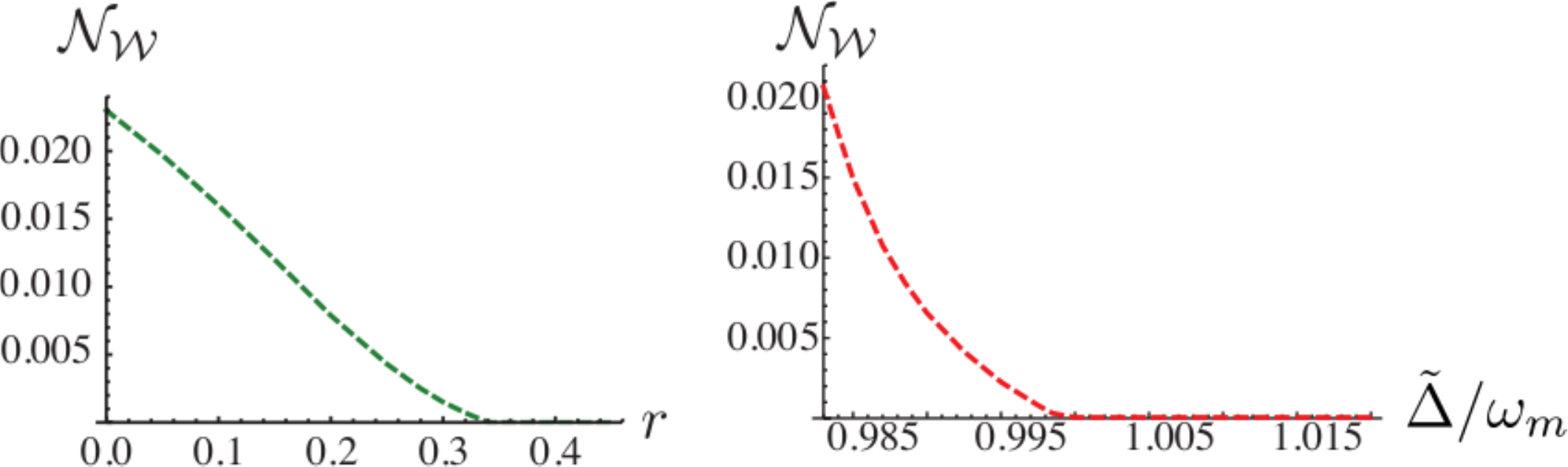}\includegraphics[width=0.70\columnwidth]{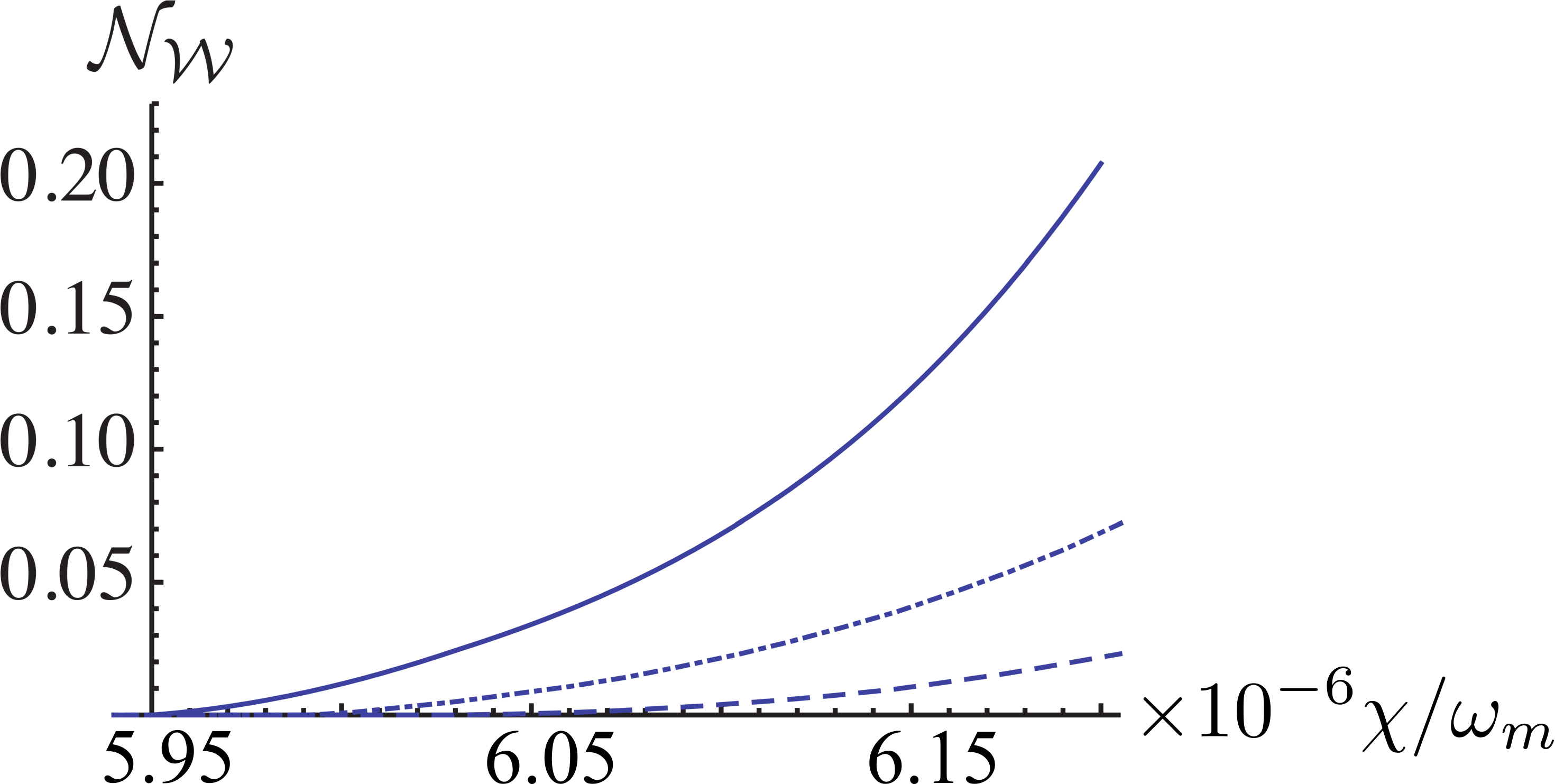}
\caption{(Color online) {\bf (a)} ${\cal N_W}$ against the squeezing parameter $r$ of the external pump ($\phi=0$). We show the results associated with the performance of Geiger-like measurements on optical field with $T=0.1$ mK, 
$\chi/\omega_m=6.2\times10^{-6}$, and $\tilde\Delta=\omega_m$. {\bf (b)} Same as in panel {\bf (a)} but shown against $\tilde\Delta$ for $T=0.1$ mK, 
$\chi/\omega_m=6.0\times10^{-6}$, $r=0$. {\bf (c)} ${\cal N_W}$ against $\chi/\omega_m$ for $T=0.1$ mK, $r=0$, $\tilde\Delta=\omega_m$. We show the results corresponding to Geiger-like detection (dashed line) and photon-counting [for $n{=}1$ (dot-dashed line) and $n{=}2$ (solid line)]. Other parameters are taken from the experiment reported in Ref.~\cite{parameters}, where a slightly lower input power was used: $\omega_m/2\pi=947$ KHz, $\gamma/\omega_m=1.5\times 10^{-4}$ with a cavity  
having wavelength $1064$ nm, decay rate $\kappa/\omega_m=0.23$ 
and pumped by a laser field of 20 mW power. However, as remarked before, similar features are observed in other experimental configurations, such as those in Ref.~\cite{Painter}.}
\label{fig4}
\end{figure*}


Armed with these tools, we are now in a position to put in place our conditional-measurement strategy. In order to do so, we rewrite the 4$\times$4 CM of the system ${\bf V}(t)$ in the block-matrix form
\begin{equation}
{\bf V}=
\begin{pmatrix}
{\bf m} &{\bf  c} \\
{\bf c}^T & {\bf f} \\
\end{pmatrix},
\label{eq5}
\end{equation}
where ${\bf m}$, ${\bf f}$ are $2\times2$ matrices accounting for the mirror and cavity-field, respectively, while ${\bf c}$ consists of the mirror-cavity correlations. The conditional mechanical state after a projective measurement over a pure Gaussian state of the field having CM ${\bf d}$ gives rise to the "updated" mechanical-mode CM~\cite{Eisert}
\begin{equation}
{\bf m}{'}={\bf m}-{\bf c}({\bf f}+{\bf d})^{-1}{\bf c}^T,
\end{equation}
independently of the measurement outcome. Homodyning is formally equivalent to projections on infinitely squeezed states, in which case 
\begin{equation}
\label{homo}
{\bf m}'={\bf m}-{\bf c}(\pi{\bf f}\pi)_{mp}{\bf c}^T
\end{equation}
with  $\pi=\text{diag}[1,0]$ and the subscript $mp$ standing for the Moore-Penrose pseudo-inverse~\cite{Eisert}. Heterodyne measurements, on the other hand, correspond to projections onto coherent states, such that ${\bm d}=\text{diag}[1,1]$. Finally, we include non-Gaussian measurements in our toolbox by considering photon-counting and avalanche photon-detectors that cannot resolve individual photons (referred to as Geiger-like detection in what follows) of the field mode. These give rise to the following conditional density matrices for the mechanical system 
\begin{equation}
\begin{split}
&\rho_m^{pc}=\text{Tr}_c\left[|n\rangle_c\langle n|\,\rho_{cm}\right],~~\rho_m^{G}=\text{Tr}_c\left[\sum_{n{=}1}^{\infty} |n\rangle_c\langle n|\,\rho_{cm}\right],
\end{split}
\label{eq12}
\end{equation}
where $\ket{n}_c$ is an $n$-photon Fock state of the cavity mode. 

\begin{figure*}[t]
{\bf (a)}\hskip4.7cm{\bf (b)}\hskip4.7cm{\bf (c)}
\includegraphics[width=0.33\linewidth]{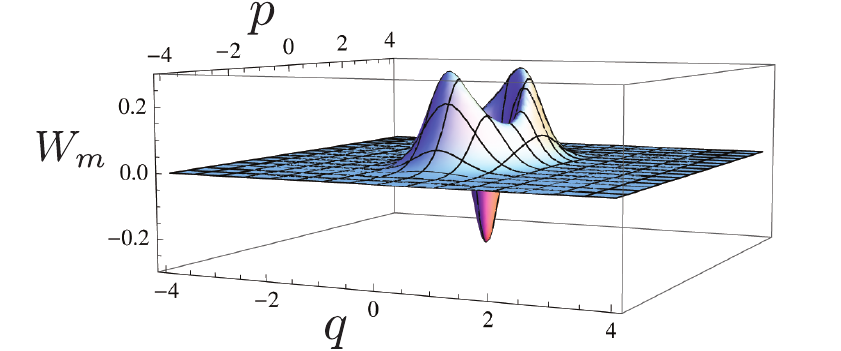}~~~\includegraphics[width=0.33\linewidth]{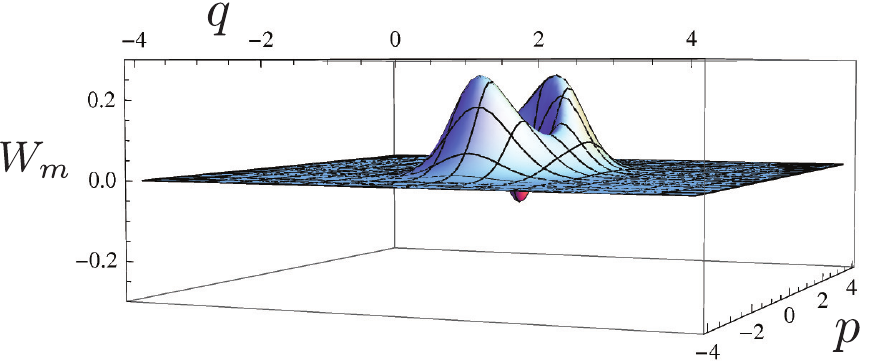}~~~\includegraphics[width=0.33\linewidth]{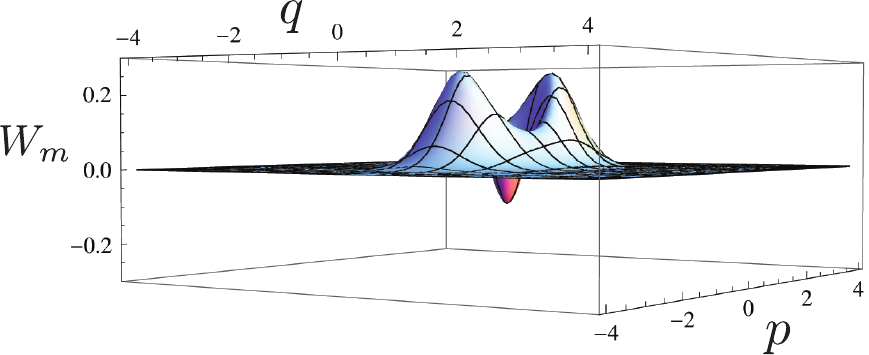}
\caption{(Color online) Wigner functions of the mechanical mode. {\bf (a)} Same conditions as in Fig.~\ref{fig4} {\bf (c)}, but we take 
$\chi/\omega_m=6.2\times10^{-6}$ with photon-counting ($n{=}1$) on the  field. {\bf (b)} Same as in panel {\bf (a)} with a temperature of $T=0.01$ K. {\bf (c)} Same as in panel {\bf (a)} but with the mechanical damping rate increased to $\gamma/\omega_m=0.015$.}
\label{fig5}
\end{figure*}

Owing to the Gaussian-preserving nature of the linearized optomechanical interaction, any conditional process based on Gaussian measurements will not generate any negativity in the Wigner function of the  mechanical mode. This contrasts explicitly the results achieved in the fully non-linear picture. Differently, both the photon-counting and the Geiger-like detection of the optical field induce strongly non-classical mechanical states having negative Wigner functions. In Fig.~\ref{fig4} we plot the negative volume ${\cal N}_{\cal W}$ of the Wigner function of the mechanical mode against some of the most relevant parameters of the model for a Geiger-like detection process, showing that non-classicality, robust against the effects of the environment affecting the optomechanical device, can indeed be induced. Quite noticeably, non-classicality at the input of the system (as embodied by a squeezed pumping field) is not a pre-requisite for the success of the process: even classical pumping will be sufficient to achieve a negative Wigner function. It is worth stressing that the negativity of the Wigner function should be taken, in general, as a sort of witness for non-classicality. However, note that non-classical states (such as multi-mode squeezed states) may well exhibit non-negative Wigner functions. 

The non-classical features of the mechanical state are robust against both an increase in the operating temperature and the mechanical damping rate. As shown in Fig.~\ref{fig5}, the Wigner function of the mechanical mode maintains its negativity as the temperature or the mechanical damping rate increase by two orders of magnitude. Some robustness is also found against the cavity decay rate: the Wigner function remains negative against an increase in $\kappa$ of about $10\%$ the value used in Figs.~\ref{fig4} and \ref{fig5}. 
Moreover, at least for projections onto low-excitation Fock states, the photon-counting strategy offers quantitative advantages over the Geiger-like one: for the same values of the optomechanical coupling rate $\chi$, ${\cal N}_{\cal W}$ associated with photon-counting is larger than the values achieved by using a Geiger-like detector, suggesting that the efficiency of the photon-counting-based mechanism is non-monotonic with $n$.

\begin{figure}[b!]
\includegraphics[width=0.6\linewidth,angle=0]{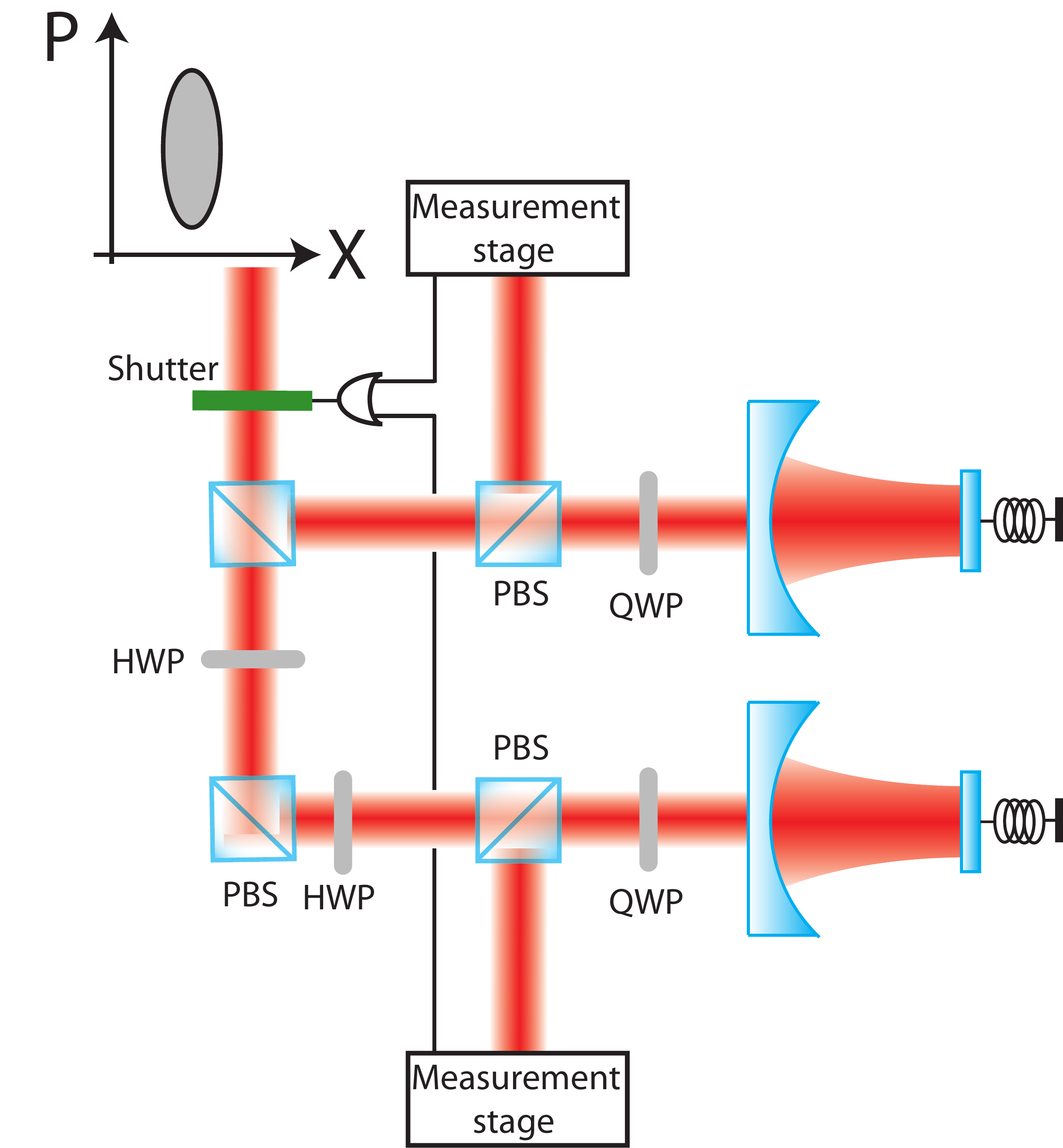}
\caption{Proposed configuration for the enhancement of purely mechanical entanglement through optical measurements. A quantum correlated state of two optical modes (generated by superimposing a single-mode squeezed state to a vacuum-state mode at a $50:50$ beam splitter) drives two remote optomechanical cavities, which are also pumped by two strong classical fields (not shown). The field leaking from the resonators is routed to two detection stages where either homodyne or heterodyne measurements are performed. {Ultra-fast electronics activates a shutter only when a measurement signal from both the detection stages is acquired (as shown by the logical OR gate)}.} 
\label{modeldouble}
\end{figure}

\section{Enhancing mechanical entanglement}
\label{ent}

We now extend the above considerations to demonstrate that all-optical post-selection can help in setting sizable all-mechanical entanglement. 

The model we consider  has been used in Refs.~\cite{Mazzola,Paternostro2012}. A sketch is illustrated in Fig.~\ref{modeldouble}, showing that it consists of two independent and noninteracting optomechanical devices, fully analogous to the one addressed in the previous Section. The total Hamiltonians of the optomechanical devices, in a frame rotating at the frequency of the lasers, thus reads
\begin{equation}
\hat{H}=\sum_{i=1,2}\hbar\Delta_i\hat{n}_i-\hbar\chi_i\hat{n}_i\hat{q}_i+\omega_m^i(\hat{p}_i^2+\hat{q}_i^2)
+i\hbar\varepsilon_i(\hat{c}_i^{\dag}-\hat{c}_i)
\end{equation}
where we have used a notation fully in line with the one introduced in the previous Section. 
The main difference with respect to the situation considered previously in this paper is that we now take the two cavities as pumped by light prepared in a pure Gaussian non-classically correlated state. As any pure entangled Gaussian state is locally equivalent to a two-mode squeezed vacuum state, we take input modes of frequency $\omega_s=\omega_L+\omega_m$ (for simplicity, we assume equal mechanical frequencies $\omega_m=\omega_m^{1,2}$) and input-noise correlations of the form
\begin{equation}
\begin{aligned}
C_j&=(\langle\delta\hat{c}_{in}^{j\dag}\delta\hat{c}_{in}^{j}\rangle, \langle\delta\hat{c}_{in}^{j}\delta\hat{c}_{in}^{j\dag}\rangle, \langle\delta\hat{c}_{in}^{j}\delta\hat{c}_{in}^{k}\rangle, \langle\delta\hat{c}_{in}^{j\dag}\delta\hat{c}_{in}^{k\dag}\rangle)\\
&=R(t,t')\delta(t-t')~~~~~(j{\ne}k{=}1,2)
\end{aligned}
\end{equation}
with $R(t,t'){=}(Z, Z{+}1, e^{-i\omega_m(t+t')}W, e^{i\omega_m(t+t')}W^*)$ and $Z=\sinh^2{r}$ and $W=\sinh{r}\cosh{r}$. The Gaussian assumption is not limiting, as this class of states are those that are routinely produced in quantum optics laboratories and two-mode squeezing is a natural way to prepare entangled field modes. Equivalently, as illustrated in Fig.~\ref{modeldouble}, a single-mode squeezed vacuum field can be superimposed to vacuum on a $50:50$ beam splitter to generate a state locally equivalent to two-mode squeezed vacuum.

The approach used in Sec.~\ref{nonclass} is now adapted to the case of the double optomechanical setting in Fig.~\ref{modeldouble}, where the light modes that have interacted with the mechanical systems enter a double-detection stage. In Ref.~\cite{Mazzola}, it has been shown that entanglement between the mechanical modes can only be achieved by means of local optomechanical interactions and thanks to the pre-available all-optical quantum correlations. In that scheme, light is simply discarded after its interaction with the mechanical modes (labelled $1$ and $2$). Our task here is to demonstrate that a significant improvement is possible by complementing the scheme with a double optical post-selection scheme, similarly to what has been illustrated in Sec.~\ref{nonclass}.

For the Gaussian-measurement scenario, we introduce the 8$\times$8 CM 
\begin{equation}
{\bm V}=
\begin{pmatrix}
{\bf M} & {\bf C} \\
{\bf C}^T & {\bf F} \\
\end{pmatrix},
\label{eq5}
\end{equation}
where ${\bf M}$, ${\bf F}$ and ${\bf C}$ are $4\times4$ block matrices accounting for the two-mirror, two-field and mirror-field properties, respectively. The entanglement shared by mechanical modes 1 and 2 is quantified using logarithmic negativity, which is defined as ${\cal L_N}=\text{max}[0, -\text{ln}2\tilde{\nu}_-]$, where $\tilde{\nu}_-=\text{min\,eig}|i\Omega_2\tilde{\sigma}|$ is the minimum symplectic eigenvalue of the matrix $\tilde{\sigma}{=}P_{1|2}\sigma_M P_{1|2}$ associated with the partially transposed state of the mechanical system (here $P_{1|2}=\text{diag}(1,1,1,-1)$, $\sigma_M$ is the CM of mechanical modes, either with or without post-selection, and $\Omega_2=\oplus^2_{j=1}i\sigma_y$ with $\sigma_y$ the $y$-Pauli matrix). We are now in a position to assess the non-classical correlations set between the two mechanical modes and how post-selection affects them. 

\begin{figure}[t]
{\bf (a)}\hskip4.5cm{\bf (b)}
\includegraphics[width=\linewidth]{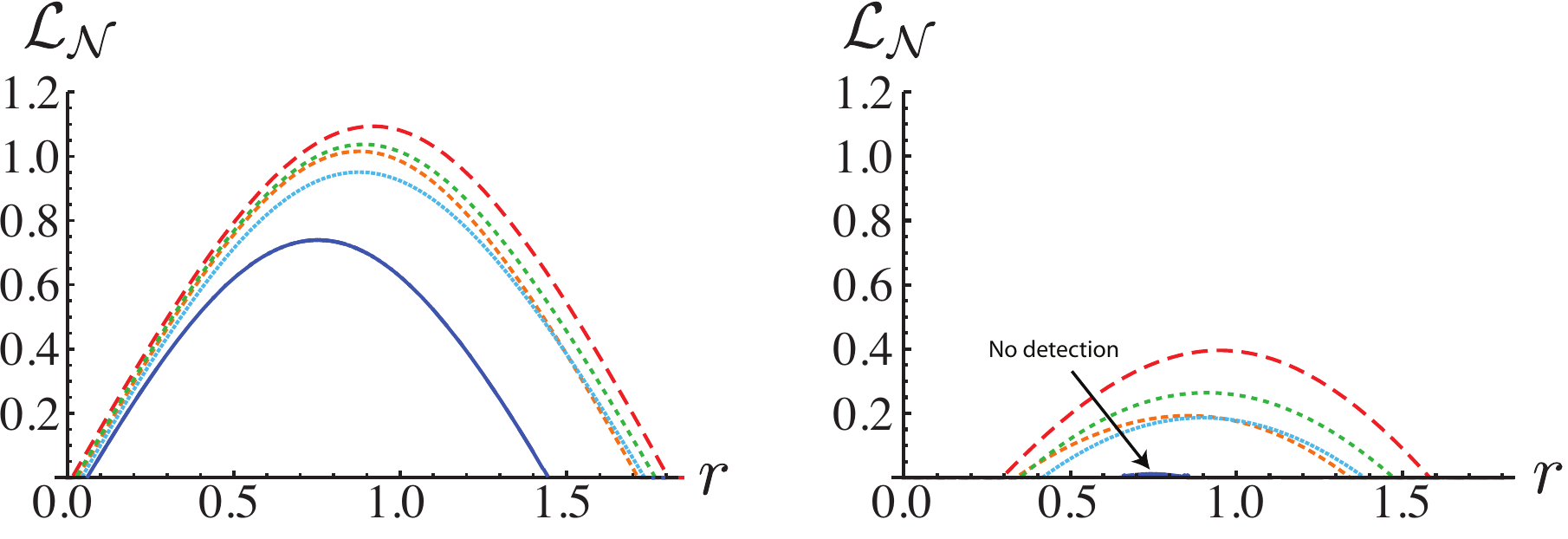}
\caption{(Color online) {\bf (a)} Mechanical entanglement ${\cal L_N}$ against the squeezing parameter $r$ of an input two-mode squeezed vacuum. We have used identical mechanical systems with $\omega^{1,2}_m/2\pi\equiv\omega_m/2\pi=947$ KHz, 
and $\gamma_m^{1,2}/\omega_m=1.5\times10^{-4}$ and cavities with 
wavelength $1064$ nm, decay rates $\kappa_{1,2}/\omega_m=0.23$ and pumped by laser fields of 20 mW power~\cite{parameters}. We have considered $\chi_{1,2}/\omega_m=3\times10^{-6}$. Curves from bottom to top: no detection; double-side heterodyne; single-side homodyne; single-side heterodyne; double-side homodyne detection. All curves are plotted 
for $\tilde\Delta_{1,2}=\omega_m$. {\bf (b)} Same as in panel {\bf (a)} but for temperature $T_{1,2}$=0.018 K.}
\label{homoheter}
\end{figure}

We perform both homodyne and heterodyne measurements as, differently from the case of single-mode non-classicality, the non-Gaussian measurements embodied by photon-counting and/or Geiger-like detections do not result in any advantage with respect to the no-measurement approach. 
 The results for Gaussian measurements are presented in Fig.~\ref{homoheter}, and include both double- and single-sided measurements (i.e. the cases where both the optical fields or a single one are detected, the latter case being independent of which mode is assessed). Without loss of generality, we have taken identical optomechanical systems. As is shown, homodyne detections indeed improve the mechanical entanglement, and detection on both cavities is superior to the measurement performed on a single one. Up to about 50\% entanglement is enhanced by homodyne detections and similar results are achieved by heterodyning, although a neat hierarchy between the two measurement strategies cannot be established. Indeed, while heterodyne detections enhance considerably the established mechanical entanglement, along the way of homodyning, the double-sided heterodyning appears to be inferior to the single-sided one, in contrast with the homodyne strategy. Further exploration of such results are currently under investigation~\cite{Jie2}.

\section{Conclusions}
\label{conc}

We have assessed the effects of post-selection on the achievement of non-classicality on a quantum mechanical mode that is driven optically by the field of an optomechanical cavity. We have demonstrated the ability of Gaussian and non-Gaussian measurements to generate mechanical states that exhibit explicitly negative Wigner functions whose non-classical features survive the effects of unfavorable working conditions. This approach has then been extended to the establishment of fully mechanical entanglement in a double optomechanical setting driven by entangled light fields. We showed that experimentally easily accessible Gaussian measurements (homodyning and heterodyning) are indeed able to boost the efficiency of such an entanglement distribution process~\cite{Mazzola,Paternostro2012} up to the achievement of larger entanglement values for a wider range of working conditions. This study contributes to the ongoing attempts to establish optomechanical analogues of current atom-light interfaces for quantum memories and networking.

\acknowledgments 
JL thanks P. L. Knight for helpful discussions on non-classicality. We acknowledge funding from the Purser Studentship, and the UK EPSRC through a Career Acceleration Fellowship and  the ``New Directions for EPSRC Research Leaders" initiative (EP/G004579/1). SG acknowledges support from the European Commission through a Marie Curie fellowship.

\renewcommand{\theequation}{A-\arabic{equation}}
\setcounter{equation}{0}
\section*{APPENDIX}  
\label{app1}

Here we provide the frequency-domain form of the elements of vector $\delta\hat{O}$ that are the roots of Eq. (7). In a compact way, the element $\delta\hat O_j~(j=1,..,4)$ of such vector of fluctuation quadratures can be cast into the form
\begin{equation}
\delta\hat{O}_j(\omega)=[A_j\delta\hat{c}_{in}(\omega)+B_j\delta\hat{c}_{in}^{\dag}(\omega)+C_j\hat{\zeta}(\omega)]/d\\
\end{equation}
with 
\begin{equation}
d=4c_s^2\chi^2\tilde\Delta\omega_m+[\tilde\Delta^2+(\kappa-i\omega)^2](i\gamma_m\omega+\omega^2-\omega_m^2).
\end{equation}
The triplets of coefficients $\tau_j=(A_j,B_j,C_j)$ are functions of the frequency $\omega$ (omitted here for easiness of notation) are then
\begin{equation}
\begin{aligned}
\tau_q&=
\begin{pmatrix}
{2c_s\chi\sqrt{\kappa}[-\kappa+i(\tilde\Delta+\omega)]\omega_m}\\
-{2c_s\chi\sqrt{\kappa}[\kappa+i(\tilde\Delta-\omega)]\omega_m}\\
{[\kappa+i(\tilde\Delta-\omega)][-\kappa+i(\tilde\Delta+\omega)]\omega_m}\\
\end{pmatrix},\\
\tau_p&=
\begin{pmatrix}
{2c_s\chi\sqrt{\kappa}(i\kappa+\tilde\Delta+\omega)\omega}\\
{2ic_s\chi\sqrt{\kappa}[\kappa+i(\tilde\Delta-\omega)]\omega}\\
{[\kappa+i(\tilde\Delta-\omega)](i\kappa+\tilde\Delta+\omega)\omega}\\
\end{pmatrix},\\
\tau_x&=
\begin{pmatrix}
{\sqrt{\kappa}(i\kappa+\tilde\Delta+\omega)[\gamma_m\omega-i(\omega^2-\omega_m^2)]}\\
{\sqrt{\kappa}(i\kappa-\tilde\Delta+\omega)[\gamma_m\omega-i(\omega^2-\omega_m^2)]}\\
-{2c_s\chi\tilde\Delta\omega_m}\\
\end{pmatrix},
\end{aligned}
\end{equation}
and
\begin{equation}
\tau_y=
\begin{pmatrix}
{-4c_s^2\chi^2\sqrt{\kappa}\omega_m+\sqrt{\kappa}(-i\kappa-\tilde\Delta-\omega)(i\gamma_m\omega+\omega^2-\omega_m^2)}\\
{-4c_s^2\chi^2\sqrt{\kappa}\omega_m+\sqrt{\kappa}(i\kappa-\tilde\Delta+\omega)(i\gamma_m\omega+\omega^2-\omega_m^2)}\\
-{2c_s\chi(\kappa-i\omega)\omega_m}\\
\end{pmatrix}
\end{equation}
This expressions are found by taking $c_s$ as real, an assumption that can be easily simplified, yet allows for considerable simplifications in the form of our solutions~\cite{mauroNJP}.

\end{document}